\documentstyle[floats,multicol,aps,epsf,prl]{revtex}

\begin{document}
\draft \twocolumn[\hsize\textwidth\columnwidth\hsize\csname
@twocolumnfalse\endcsname
\def\btt#1{{\tt$\backslash$#1}}
\title{Oscillatory dynamics and organization of the vortex solid in YBa$_2$Cu$_3$O$_7$ single crystals}
\author{ S. O. Valenzuela and V. Bekeris}
\address{Laboratorio de Bajas Temperaturas, Departamento de
F\'{\i}sica, Universidad Nacional de Buenos Aires, Pabell\'on I, Ciudad
Universitaria, 1428 Buenos Aires, Argentina}

\maketitle
\begin{abstract}
{ We report on the degree of order of the vortex solid in
YBa$_2$Cu$_3$O$_7$ single crystals observed in ac susceptibility
measurements. We show that when vortices are ``shaken" by a
temporarily symmetric ac field they are driven into an
easy-to-move, ordered structure but, on the contrary, when the ac
field is temporarily asymmetric, they are driven into a more
pinned disordered state. This is characteristic of tearing of the
vortex lattice and shows that ordering due to symmetric ac fields
is essentially different from an equilibration process or a
dynamical crystallization that is expected to occur at high
driving currents.}

\end{abstract}
%\baselineskip=24 pt
%\draft
]
%\narrowtext

Vortex lattices (VL) in type II superconductors provide a unique
system with tuneable parameters for exploring driven dynamic
phases. A characteristic feature of driven dynamics in quenched
disorder is the existence of a depinning transition. Experimental
\cite{thor} and theoretical \cite{kosh} results have shown
evidence of a two-step process in the depinning transition of the
VL as the driving force is increased: first it undergoes plastic
flow, where neighboring parts of it move at different velocities,
and then, above a threshold force $F_T$, a dynamic crystallization
occurs as proposed by Koshelev and Vinokur \cite{kosh} (K-V) where
all the vortices move at the same average velocity \cite{pld}.
With the use of fast current ramps and ac techniques \cite{hend},
the attention has been focused on the evolution of the VL as it
depins and starts moving. Though the idea of this annealing
process is accepted in steady driven VL's, a number of new
puzzling phenomena were observed with these techniques that cannot
be ascribed to it such as complex history dependent dynamic
response and memory effects \cite{hend,nos,pal}. To explain these
phenomena, it has been proposed \cite{pal} that the overall vortex
structure surges as a balance between the injection of disordered
vortices through surface barriers and an annealing by the driving
current. However, this seems not to be the case in the high $T_c$
twinned YBa$_2$Cu$_3$O$_7$ (YBCO) single crystals, where similar
effects have been observed, but only bulk pinning appears to be
relevant \cite{nos}. In the latter investigation, it was found
that an ``easy- to-move'' more ordered VL can be stabilized by the
shaking movement induced by a temporarily symmetric ac field ({\it
e.g.} sinusoidal). Very recently, it has been suggested that this
reordering may be a consequence of a current-assisted transition
from a supercooled metastable phase to the stable equilibrium
phase \cite{xiao,chad}. The fluctuation energy produced in the
sample under cyclic field variation would be a key element within
this scenario \cite{chad}.

In this Letter, we show that if the shaking field is temporarily
{\it asymmetric} ({\it e.g.} sawtooth) a disordered VL can be
recovered. The ``hard-to-move'' state so obtained would have a
high density of topological defects due to the plastic flow of
vortex bundles sliding one past another. The fact that the degree
of order is strongly dependent on the temporal symmetry of the ac
field would indicate that the reordering observed in Ref.
\cite{nos} is due neither to a K-V transition nor to a
current-assisted equilibration but to a process of a fundamentally
different kind which involves repeated interaction of neighboring
vortices.

We also propose that when the VL is pushed in and out of the
sample with an asymmetric ac field, it is torn up in a
``ratchet-like'' fashion as a consequence of the different Lorentz
forces involved in the ramp up and ramp down of the field.
Important related results show that these plastic distortions are
partly reversible when the temporal evolution of the ac field is
reversed.

The results shown here are for a single crystal sample of YBCO
\cite{ale} (dimensions $0.56 \times 0.6 \times 0.02 ~
\mathrm{mm}^{3}$) with $T_{\mathrm{c}}$ = 92 K and $\delta T_c$ =
0.3 K determined by ac susceptibility ($h_{ac}$=1 Oe) at zero dc
field. The same sample was used in previous studies on plasticity
and memory effects \cite{nos}. Measurements were done with the
standard mutual inductance technique, with the ac field parallel
to the $c$ axis of the sample. The in-phase and out-of-phase
($90^{\circ}$) signals were collected with a lock-in amplifier.
Data were recorded for an angle $\theta = 20^{\circ}$ between the
dc applied field, $H_{dc}$, and the $c$ axis to avoid the
Bose-glass phase \cite{nos}. The dc field was oriented out of all
twin boundaries simultaneously (see inset in Fig. 1).

Fig. 1 presents ac susceptibility measurements that correspond to
different thermomagnetic histories of the sample. Measurements
were performed by varying $T$ (in the direction indicated by the
arrows) at a rate of 0.2 K/min, at fixed $H_{\mathrm{dc}}$ (3
kOe), $h_{\mathrm{ac}}$ (1.6 Oe) and frequency (10.22 kHz)
(default parameters). Dashed curves were obtained in the usual
field cooled procedure (F$_{\mathrm{ac}}$C C) while solid ones
were obtained on warming after zero (ac and dc) field cooling (ZFC
W) (the ac field is turned on after the dc field has reached its
final value \cite{com1}). We believe that the ZFC W case
corresponds to a disordered state because of the plastic motion of
vortices as they penetrate into the sample \cite{nos}. In
addition, in the F$_{\mathrm{ac}}$C C case, the VL is partially
annealed by the shaking movement induced by the ac field during
the cooling process \cite{pan}. As discussed previously
\cite{nos}, the history effects are observed because of a change
in the mobility of the VL and the fact that the inner disordered
state in the ZFC W case can sustain a higher current without
vortex movement thereby enhancing the shielding $|\chi'|$ and
reducing the dissipation $\chi$'' (for small measuring ac fields).

\begin{figure}[btp]

\epsfxsize=3.2in

\vskip -5mm

\epsfbox{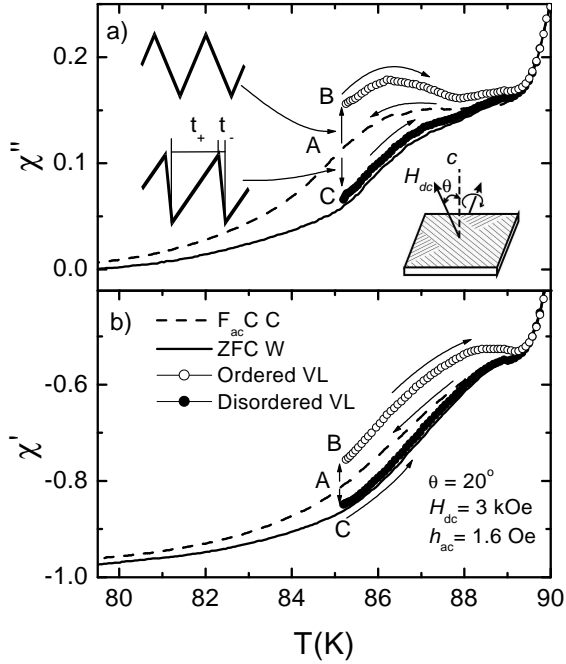}

\vskip -11mm

\protect\caption{$\chi''(T)$ (a) and $\chi'(T)$ (b) for the YBCO
single crystal with different thermomagnetic histories. Arrows
indicate the direction of the temperature sweep. Dashed curves
were obtained on cooling, F$_{ac}$C C. Solid curves were obtained
on warming after cooling with neither ac nor dc field applied, ZFC
W. Application of $10^5$ cycles of a triangular ac field (6.5 Oe,
10 kHz) to a F$_{ac}$C C state at 85.2K (A) leads to B and upper
curves (open circles). Application of $10^5$ cycles of a sawtooth
ac field (6.5 Oe, 10 kHz) to a F$_{ac}$C C state at 85.2K (A)
leads to C and lower curves (solid circles). In all cases $h_{ac}$
$\parallel c$ = 1.6 Oe and $H_{dc}$ = 3 kOe is applied at $\theta
= 20 ^{\circ}$ from the {\it c} axis (see lower inset in the upper
panel.)} \label{fig1}
\end{figure}

Through the F$_{\mathrm{ac}}$C C procedure one can obtain the most
ordered state attainable with the measuring ac field. However, a
more ordered VL can be obtained if the VL is shaken more intensely
with a higher ac field. To illustrate this, we prepared a
F$_{\mathrm{ac}}$C C state at 85.2 K (point A) and applied for a
short time (ten seconds) a triangular ac field with $h_{ac}$ = 6.5
Oe and $f$= 10 kHz. We turned the 6.5 Oe ac field off and measured
the ac susceptibility with our default parameters on warming. The
ordering of the VL is inferred from the unambiguous changes of
$\chi'$ and $\chi''$ (point B and upper curves in Fig. 1, open
circles).

Our first important result surges when contrasting these
measurements with the ones obtained after the application of an
{\it asymmetric} ac field under the same conditions. We prepared
again a F$_{\mathrm{ac}}$C C state at 85.2 K (point A) and applied
during ten seconds a sawtooth ac magnetic field ($h_{ac}$ = 6.5
Oe, $f$ = 10 kHz). The magnetic field waveform has a rising time,
$t_+$, and a shorter falling time, $t_-$, (see inset in Fig. 1(a))
so that a given current forces vortices, for example, towards the
center of the sample during the longer time interval, but a higher
current forces them out during the rest of the cycle. We turned
the 6.5 Oe ac field off and then measured the ac susceptibility.
The result is indicated as point C in the lower curves of Fig.
1(a) and (b), where it is clearly seen that $|\chi'|$ increases
and $\chi''$ decreases. On further warming, we found that the
measured susceptibility (solid circles) goes close to the {\it
disordered} ZFC W curve. The difference in the response after the
application of a symmetric or an asymmetric ac field is striking.
The sawtooth ac field has brought the VL to a more pinned
disordered state. It is worth noting here that one can go from
point C to point B, or vice versa, by properly applying a high
amplitude symmetric or asymmetric ac field as explained above.

Fig. 2 exhibits clear evidence that both the ordering and
disordering of the VL sketched in Fig. 1 are number-of-cycles
dependent. The same information can be extracted from $\chi'$ and
$\chi''$. As the relative change in $\chi''$ is larger than in
$\chi'$, we only present $\chi''$ data. The initial states are
obtained after exposing the sample to $10^4$ cycles of sawtooth ac
field (initial disordered state in Fig 2(a)), plus $10^5$ of
triangular ac field (initial ordered state in Fig. 2(b))
\cite{com2}.

Fig. 2(a) shows the dynamics of the VL's ordering. Starting from a
disordered hard-to-move state, the sample is cycled with a
symmetrical (sinusoidal, triangular or square) ac field ($h_{ac}$=
6.5 Oe, f= 10 kHz). After $N_s$ cycles, the ac field is turned off
and the state of the VL is inspected by measuring $\chi_{ac}$ with
a smaller probe ac field (default parameters above). Before
repeating the procedure for a new value of $N_s$, the VL is
brought again to the initial disordered state. It is seen that the
dependence of $\chi_{ac}$ on $N_s$ is approximately logarithmic,
and saturation, if any, occurs beyond $10^6$ cycles.

\begin{figure}[btp]

\epsfxsize=3.3in

\vskip -5mm

\epsfbox{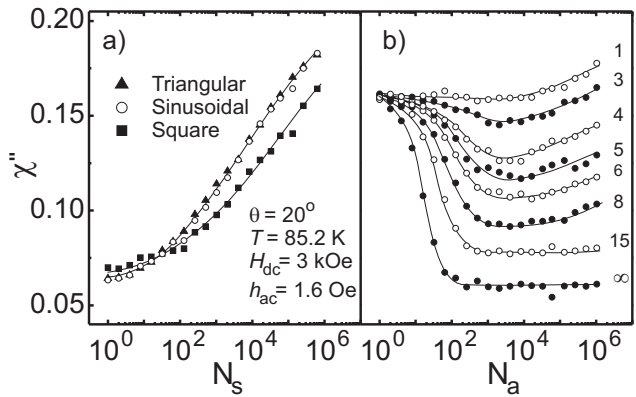}

\vskip -5mm

\protect\caption{ Oscillatory organization of vortices. (a)
Ordering of an initially disordered VL: $\chi''$ vs. $N_s$ for
triangular (solid triangles), sinusoidal (open circles) and square
(solid squares) ac field symmetric waveforms with $h_{ac}$ = 6.5
Oe and $f$ = 10 kHz. (b) Disordering of an initially ordered VL:
$\chi''$ vs. $N_a$ for different asymmetries $\alpha$ of the field
(see text), $h_{ac}$ = 6.5 Oe and $f$ = 10 kHz. $\alpha$ = 1
corresponds to a triangular waveform  and $\alpha$ = $\infty$ to a
sawtooth waveform. Measuring ac field: 1.6 Oe, 10.22 kHz. Lines
are guides to the eye.}

\label{fig2}
\end{figure}

The evolution of the disordering was inspected in a similar
experiment at the same temperature. In Fig. 2(b) we show
$\chi_{ac}$ as a function of $N_a$ asymmetric ac field cycles
applied to an initially easy-to-move ordered lattice. The
asymmetry of the waveform is defined as $\alpha$ = $(t_+ / t_-)$.
In all cases, the preparation of the initial ordered state before
the application of the next $N_a$ asymmetric ac field cycles
erases any eventual effects of the small ac probe used in the
$\chi_{ac}$ measurement.

Ordering due to the application of a driven current or an
oscillatory perturbation has been frequently observed and proposed
to explain some unusual behaviors \cite{hend,nos,pal,xiao,chad}.
Some of these results have been interpreted as an equilibration
process \cite{xiao,chad} or as a combined effect of the injection
of disordered vortices through the sample edges and a
crystallization assisted by the driving current \cite{pal}. In the
former case, it has been proposed \cite{chad} that metastable to
stable transformations occur in the regions of the sample where
the local energy dissipation exceeds a threshold value.

In our experiments, it is clear that an asymmetric ac field will
also produce an oscillatory perturbation; but then it is
legitimate to inquire into why the VL disorders. The discrepancy
that this implies suggests that the ordering process is unlikely
to be a K-V crystallization and that it is surely not an
equilibration transition where energy dissipation in the sample
will always tend to order the VL \cite{chad}.

In the K-V scenario \cite{kosh}, ordering is only achieved when
the Lorentz force exceeds a certain threshold value, $F_T$, while
a plastic disordered motion occurs for forces below it. The
induced current density in our experiments depends directly on the
sweep rate of the applied ac field which, on the other hand, is
proportional to its frequency. Dynamical reordering is observable,
at least,  for frequencies  in the range of 0.1 Hz to 300 kHz for
sinusoidal, triangular and even for the high sweep rate square
waveform. Disordering shown in Fig. 2(b) is clearly discernible
for a frequency (10 kHz) well within the range of frequencies in
which ordering is achieved and for asymmetries as small as 3. This
observation would indicate that for {\it equal values of current
density} circulating in the sample one can either order or
disorder the VL depending {\it only} on the asymmetry of the ac
field and that, remarkably, ordering can be achieved with {\it
lower} current densities ({\it e.g.}, at 0.1 Hz) than those that
are able to disorder the VL as shown in Fig. 2(b). On account of
this, the K-V transition can be discarded as a possible
explanation of our observations. In addition, the dynamically
generated disorder with the asymmetric ac field is distinctive of
tearing of the VL, which suggests that the driving force is below
$F_T$ and that the flow of vortices is plastic (first step in the
depinning process \cite{kosh,pld}). In these conditions, there are
regions of the VL where the strain is large enough to cause phase
slips and the number of vortices that participate in the movement
strongly depends on the magnitude of the applied force, {\it ie}.
on the induced current.

If the ac field is asymmetric, the vortices that move when the
field is ramped up will not be the same ones that move when the
field is ramped down. Then, after the application of one cycle of
an asymmetric ac field, there should be a net displacement between
vicinal bundles of vortices. This type of movement will cause
lattice distortions that destroys the long range order of the VL
and leads to the increase of $|\chi'|$ and the decrease of
$\chi''$ for small ac fields. Note that these distortions are
cumulative and increase with the number of cycles leading to a
rapid saturation of $\chi _{ac}$ after $\sim 300$ cycles for the
sawtooth waveform. When $\alpha$ is decreased, the disorder also
decreases and the number of cycles needed to reach maximum
disorder increases.

Recovering an ordered VL requires a larger number of cycles of a
symmetric ac field of the same amplitude. As seen in Fig. 2(a),
after $10^6$ cycles $\chi'$ and $\chi''$ still vary approximately
as  $\ln N$. It is the overall symmetry of the ac field that
drives vortices back and forth, with no net displacement between
them. The effect of this shaking movement of vortices is to
locally order the VL. Apparently, repeated interactions of
vortices with their nearest neighbors seems to enhance ordering
and favor the healing of defects.

A curious behavior is observed when {$\alpha$} is reduced: the
change in $\chi''$ is non-monotonous in the number of cycles. This
is more clearly seen for the smaller asymmetries. Initially,
$\chi''$ rapidly decreases but after $\sim 10^4$ cycles this
tendency is reversed and $\chi''$ slowly starts to increase. This
phenomenon could be explained through a process involving dynamic
and static friction. The argument could be as follows. At the
beginning, the VL disorders as described above as a consequence of
the asymmetry of the ac field. However, as $\alpha$ is small, the
bundles of vortices that {\it just} start moving when subjected to
the larger Lorentz force (higher field sweep rate), were at the
static limit in the preceding part of the field cycle (lower field
sweep rate). Once they move, it is established a new state
determined by dynamic friction (which presumably is smaller than
the static one) in which these bundles of vortices move coherently
with the rest. If this picture is correct, it would take about
$\sim 10^4$ cycles to reach a steady state in which vortices move
coherently and the repeated interaction between them, that allows
ordering, dominates in what $\chi_{ac}$ is concerned.

\begin{figure}[btp]

\epsfxsize=3.5in

\vskip -3mm

\epsfbox{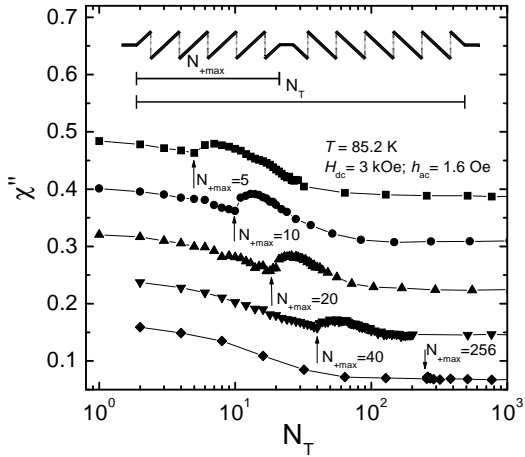}

\vskip -1mm

\protect \caption{Releasing distortions in the VL. $\chi''$ as a
function of the number $N_+$ of sawtooth field cycles applied to
an initially ordered VL. After applying $N_{+max}$ cycles, the
polarity is inverted as shown in the inset. $N_T$ is the total
number of cycles $N_+ + N-$. Curves are vertically displaced for
clarity. Measuring ac field: 1.6 Oe, 10.22 kHz. Lines are guides
to the eye.}

\label{fig3}
\end{figure}

Finally, we address the question whether the distortions produced
by the asymmetric ac field can be released if we make a temporal
inversion of the ac field. If vortex movement were completely
reversible and we applied $N_+$ cycles of an asymmetric ac field,
we would recover the initial configuration of the VL after
applying the same number of cycles of an asymmetric ac field with
the polarity inverted. The results in Fig. 3 suggest that a more
ordered configuration can be recovered whenever the distortions
generated are below certain level. As shown in the inset of Fig.
3, we first apply a given number of cycles, $N_+$, of sawtooth ac
field to an initially ordered VL. As in Fig. 2, $\chi''$ decreases
as $N_+$ increases. After $N_{+max}$ cycles, we invert the
polarity of the ac field so that the next cycles represent a
``temporal inversion'' of the previously applied ac field. At this
point it is observed that $\chi''$ starts increasing and, when the
number of cycles of the inverted waveform, $N_-$, exceeds
$N_{+max}$, $\chi''$ decreases again. We have followed this
procedure for $N_{+max}$ = 5, 10, 20, 40 and 256. When $N_{+max}$
is high enough to highly distort the VL and saturation of $\chi''$
is reached ($N_{+max}$ = 256) almost no reversibility is observed.
These results further support the scenario proposed of
ratchet-like tearing of the VL by a temporarily asymmetric ac
field.

To conclude, we have presented susceptibility measurements in a
twinned YBCO single crystal after shaking the VL with an ac field.
We find a dramatic change of the VL response when the asymmetry of
the shaking ac field is increased. While a symmetric ac field
orders the VL an asymmetric ac field produces the opposite effect.
The experiments described demonstrate that ordering is not due to
a K-V transition or to an equilibration process. The ordering of
the VL appears to be a consequence of the repeated interaction of
neighboring vortices while the disorder produced by an asymmetric
ac field would be related to tearing caused by the different
Lorentz forces involved in the ramp up and down of the ac field.
We hope these results will stimulate both experimental and
theoretical work on the dynamics of VL's under an oscillatory
perturbation to further test the hypotheses presented here.

We acknowledge E. Osquiguil for a critical reading of the
manuscript. This research was partially supported by UBACyT TX-90,
CONICET PID N$^{\circ}$ 4634 and Fundaci\'on Sauber\'an.


\begin{references}
\bibitem{com1} Note that the dc field is more than two orders of magnitude higher than the
full penetration field at the temperatures of interest. See Ref.
5.
\bibitem{com2}We have chosen this procedure because the saturation value of
$\chi_{ac}$ reached after the application of a sawtooth ac field
is found to be independent of the history of the sample.
\end{references}
\end{document}